\documentclass[a4paper,12pt]{article}
\usepackage[centertags]{amsmath}
\usepackage{amsfonts}
\usepackage{amssymb}
\usepackage{amsthm}
\usepackage{newlfont}
\usepackage{mathrsfs}
\usepackage{pst-node}
\usepackage{palatino}  
\usepackage{fancyhdr}
\usepackage{graphicx}
\usepackage{fancybox}
\usepackage{geometry}
\usepackage[dvips]{epsfig}

  \let\La=\Lambda

\newcommand{\bbZ}{{\mathbb Z}}

\newcommand{\opunit}{\text{1}\kern-0.22em\text{l}}

\newcommand{\bde}{\begin{definition}}
\newcommand{\ede}{\end{definition}}
\newcommand{\beq}{\begin{equation}}
\newcommand{\eeq}{\end{equation}}
\newcommand{\ben}{\begin{enumerate}}
\newcommand{\een}{\end{enumerate}}
\newcommand{\ble}{\begin{lemma}}
\newcommand{\ele}{\end{lemma}}
\newcommand{\bpr}{\begin{proof}}
\newcommand{\epr}{\end{proof}}

\title{First-order transitions \\for some generalized XY  models}

\author{
  {\normalsize Aernout C.~D.~van Enter}        \\[-1mm]
  {\normalsize\it Centre for Theoretical Physics}   \\[-1.5mm]
  {\normalsize\it Rijksuniversiteit Groningen}         \\[-1.5mm]
  {\normalsize\it Nijenborgh 4}                \\[-1.5mm]
  {\normalsize\it 9747 AG Groningen}           \\[-1.5mm]
  {\normalsize\it THE NETHERLANDS}             \\[-1mm]
  {\normalsize\tt aenter@phys.rug.nl}        \\[-1mm]
\\ [-1mm]
  {\normalsize Silvano Romano}            \\[-1mm]
  {\normalsize\it Unit\`a di Ricerca CNISM  e Dipartimento di Fisica ``A.Volta''}   \\[-1.5mm]
{\normalsize\it Universit\`a di Pavia}         \\[-1.5mm]
  {\normalsize\it via A. Bassi 6}             \\[-1.5 mm]
  {\normalsize\it I-27100, Pavia}           \\[-1.5mm]
  {\normalsize\it ITALY}             \\[-1mm]
  {\normalsize\tt Silvano.Romano@pv.infn.it}        \\[-1mm]
\\ [-1mm]
  {\normalsize Valentin A. Zagrebnov}            \\[-1mm]
  {\normalsize\it Universit\'{e} de la M\'{e}diterran\'{e}e}   \\[-1.5mm]
  {\normalsize\it Centre de  Physique Th\'{e}orique}         \\[-1.5mm]
  {\normalsize\it Luminy - Case 907}                \\[-1.5mm]
  {\normalsize\it F-13288, Marseille, Cedex 9}           \\[-1.5mm]
  {\normalsize\it FRANCE}             \\[-1mm]
  {\normalsize\tt zagrebnov@cpt.univ-mrs.fr}        \\[-1mm]
{\protect\makebox[5in]{\quad}}}
\begin{document}
\maketitle \baselineskip=14pt \noindent {\bf Abstract.}
In this note  we demonstrate the occurrence of first-order transitions in
temperature for some  recently introduced generalized XY models, and also point
out the connection between them and annealed site-diluted
{(lattice-gas)} continuous-spin models.

\bigskip


\newtheorem{theorem}{Theorem}          
\newtheorem{lemma}{Lemma}              
\newtheorem{proposition}[lemma]{Proposition}
\newtheorem{corollary}[lemma]{Corollary}
\newtheorem{definition}[theorem]{Definition}
\newtheorem{conjecture}[theorem]{Conjecture}
\newtheorem{claim}[theorem]{Claim}
\newtheorem{observation}[theorem]{Observation}
\def\proof{\par\noindent{\it Proof.\ }}
\def\reff#1{(\ref{#1})}

\let\zed=\bbbz 
\let\szed=\bbbz 
\let\IR=\bbbr 
\let\R=\bbbr 
\let\sIR=\bbbr 
\let\IN=\bbbn 
\let\IC=\bbbc 

\def\nl{\medskip\par\noindent}

\def\scrb{{\cal B}}
\def\scrg{{\cal G}}
\def\scrf{{\cal F}}
\def\scrl{{\cal L}}
\def\scrr{{\cal R}}
\def\scrt{{\cal T}}
\def\pfin{{\cal S}}
\def\prob{M_{+1}}
\def\cql{C_{\rm ql}}
\def\bydef{\stackrel{\rm def}{=}}   
\def\qed{\hbox{\hskip 1cm\vrule width6pt height7pt depth1pt \hskip1pt}\bigskip}
\def\remark{\medskip\par\noindent{\bf Remark:}}
\def\remarks{\medskip\par\noindent{\bf Remarks:}}
\def\example{\medskip\par\noindent{\bf Example:}}
\def\examples{\medskip\par\noindent{\bf Examples:}}
\def\nonexamples{\medskip\par\noindent{\bf Non-examples:}}

\newenvironment{scarray}{
          \textfont0=\scriptfont0
          \scriptfont0=\scriptscriptfont0
          \textfont1=\scriptfont1
          \scriptfont1=\scriptscriptfont1
          \textfont2=\scriptfont2
          \scriptfont2=\scriptscriptfont2
          \textfont3=\scriptfont3
          \scriptfont3=\scriptscriptfont3
        \renewcommand{\arraystretch}{0.7}
        \begin{array}{c}}{\end{array}}

\def\wspec{w'_{\rm special}}
\def\mup{\widehat\mu^+}
\def\mupm{\widehat\mu^{+|-_\Lambda}}
\def\pip{\widehat\pi^+}
\def\pipm{\widehat\pi^{+|-_\La\bibitem{mi}

mbda}}
\def\ind{{\rm I}}
\def\const{{\rm const}}

\bibliographystyle{plain}


\newpage

\maketitle
\section{Introduction}
In some recent papers \cite{RZ,CRMP,MPMM,MPCR} a class of
\textit{generalized} XY models was introduced and studied. These
models are ferromagnetic and, in the simplest case, restricted to a
nearest-neighbour XY - type interaction. In contrast to the
\textit{plane rotator}
interaction, they involve $3$-component spins (just
as the classical Heisenberg model does) and they possess an $O(2)$
symmetry with respect to the X and Y components. This part is
multiplied by a product of single-site terms, depending on the third
(Z) component only, which is raised to an exponent $p > 0$. In the
following, this factor will  be simply called ``single-site term''.

{For $p=1$ one recovers the standard ferromagnetic XY model, in
which the interaction is defined by the scalar product of the
nearest neighbour spin projections on the XY - plane.}

{With the help of some correlation inequalities} it was found
{\cite{RZ}} that in $d=2$ there is a transition between a
low-temperature Berezinski\v\i-Kosterlitz-Thouless (BKT) phase and a
high-temperature phase, whereas, in $d \ge 3$  dimensions, the
existence of  magnetic order at low temperatures was established.

The nature of this transition, however, was left open in these
works. In $d=2$ one might expect either a BKT scenario (an
infinite-order transition between the BKT phase and a high-temperature phase),
which was found when this exponent $p$
is  small, or a first-order one between the BKT phase and the high-temperature
phase , and in $d=3$ there may be  an
ordinary second-order transition (again found for small $p$), or a
first-order transition. Recall that the first-order transition means
coexistence of different infinite-volume Gibbs measures. Here this
implies a jump in the energy density, in $d \geq 3$ (but not in
$d=2$) accompanied by a jump in the magnetisation.

\smallskip

In this {note} we point out that, by  a minor adaptation of the
reflection positive chessboard-estimates analysis  for free-energy
contours (which goes back to \cite{DS1, KS} and which was recently
applied to study magnetic transitions for \textit{nonlinear}
classical vector spin models in \cite{ES1,ES2,ES3,BiKo,MeNa}), the
occurrence of a first-order transition for sufficiently
\textit{large} values of the exponent $p$ can be proven. Such a
transition  was already suspected to occur in $d=2$, based on the
numerical data of \cite{MPCR}. In the last part of our paper we
compare this result with further numerical data.

\smallskip

We  remark that with some small modifications our analysis covers
also the case of the first-order, instead of e.g. a BKT
infinite-order, transition in some \textit{annealed} diluted
lattice-gas models (see \cite {CKS}, \cite{GZ}, Sect. 2.4, and
\cite{CSZ}), and even in some \textit{continuum} magnetic systems
\cite{GTZ}. In this case the role of the nonlinearity is played by
terms involving either the lattice-gas particle occupation numbers
\cite {CKS, GZ, CSZ}, or the particle density \cite{GTZ}. This
becomes in particular evident, when we consider the model of
\cite{RZ} in the \textit{square-ditch} approximation, see
(\ref{ditch-Ham}). Then the generalized XY model reduces to an
annealed {site-diluted }
\textit{plane rotator}
model.

With minor modifications we can also treat generalized Heisenberg
interactions for n-vector spins with $n \geq 3$ or annealed dilute
Heisenberg models. In that case we expect that in $d=2$ there will
be a transition between two phases with exponentially decaying
correlations, similarly to  what is expected for the nonlinear
models of \cite{ES1,ES2}.

\section{Model, proofs and results}

For  general background on the theory of Gibbs measures on lattice systems
we refer to \cite{EFS, Geo, GeoHagMae, Sin, Sim}. The method of
reflection positivity  and chessboard estimates \cite{RP}
is reviewed in \cite{Shl1} and in the last 4 chapters of \cite{Geo}. The fact
that our models  satisfy the property of reflection positivity
follows immediately from the conditions described there.

\smallskip

Our systems  are as follows. On each site of the lattice $\bbZ^d$ we
have a three-component unit spin, described by spherical coordinates
$\phi$ and $\theta$; our models are then described by {a nearest
neighbour generalized}  XY interaction, {i.e. the plane rotator
interaction in the $\phi$-variables combined with a product of
$p$-powers of} single-site terms in the $\theta$-variables
\cite{RZ}. For a finite $\Lambda \subset \bbZ^d$ the (dimensionless)
Hamiltonian reads:
\begin{equation}\label{Ham}
H^{\Lambda}(\phi, \theta): = - \sum_{\langle i,j \rangle \subset
\Lambda} [\sin (\theta_i) \sin (\theta_j)]^{p} \cos (\phi_i - \phi_j)
\end{equation}
The integer exponent $p > 0$ is a parameter in our model, and a
large value of  $p$  means that spins can only interact noticeably
when they are in a narrow \textit{ditch} around the equator $\theta =
{\pi}/{2}$, whose width is of order {$O({1}/{\sqrt p})$}.
This narrow and deep ditch plays a similar role as the narrow-well
potentials of \cite{ES1,ES2,BiKo}.

We present our proof for the \textit{two-dimensional} model where the ditch
has a \textit{square}, instead of a polynomial, shape. Extensions to
polynomial shapes then can be done as in \cite{ES1, ES2,BiKo}.
Indeed, our proof can be seen to be almost a corollary of these
papers, to which we refer for further details.

So, we consider the \textit{square-ditch} {approximation of the
generalized} XY model (\ref{Ham}) with Hamiltonians
\begin{equation}\label{ditch-Ham}
H^{\Lambda}_{\varepsilon}(\phi, \theta) := - \sum_{\langle i,j \rangle \subset
\Lambda}   \ n(\theta_i) n(\theta_j) \
\cos (\phi_i - \phi_j) \ , \  \ n(\theta):= 1_{\varepsilon} (\theta) \ ,
\end{equation}
where  $1_{\varepsilon}(x)$ denotes, for
$0\leq\varepsilon\leq\pi/2$, the characteristic function of the
interval $[\pi/2 -\varepsilon, \pi/2 +\varepsilon]$.

{The square-ditch approximation (\ref{ditch-Ham}) implies that two spins
can interact only when they both are in the ditch; in other words,
one can interpret (\ref{ditch-Ham}) as the Hamiltonian of annealed site-diluted plane rotator
model with lattice-gas occupation numbers $n = {0, 1}$, as discussed in \cite{CKS,CSZ},
and as is also suggested by the notation.

Notice that the (\textit{a priori}) one-site distribution on those occupation
numbers is induced by the uniform probability (\textit{Haar})
measure $\mu_{0}(d \phi, d \theta):= d \phi \ d \theta \ sin \theta
/4\pi$ on the unit sphere. Therefore, $\varepsilon$ is related to
the chemical potential $\nu$ that governs the lattice-gas overall
particle density. By the standard definition of the lattice-gas
chemical potential
the relation becomes:
\begin{equation}\label{chem-poten}
\nu = \beta^{-1}\ln \frac{\sin \varepsilon}{1 - \sin \varepsilon} \
\ \ ,
\end{equation}
where $\beta^{-1}=\Theta$ denotes the (dimensionless) temperature of
the system. Hence at fixed temperature the chemical  potential
becomes negative and large in magnitude when $\varepsilon$ is close
to zero, which corresponds to a small lattice-gas density; whereas
for $\varepsilon \rightarrow \pi/2$, i.e. for $\nu \rightarrow
+\infty$, one obtains a non-diluted plane rotator model
(\ref{ditch-Ham}) with  $n(\theta_j)= 1$ for all $\{\theta_j\}_j$. }

{For the proof} we consider in a two-dimensional lattice a square
$\Lambda$, of a linear size $N$ which is a multiple of $4$, with
\textit{periodic} boundary conditions. Associated to  Hamiltonians
$H^{\Lambda}_{\varepsilon}(\phi, \theta)$ are Gibbs measures
\begin{equation*}
\mu^{\Lambda}(d \phi, d \theta)= {\frac{1}{Z^{\Lambda}}} \exp
[-\beta H^{\Lambda}_{\varepsilon} (\phi, \theta)]\mu_{0}^{\Lambda}(d
\phi, d \theta) \ ,
\end{equation*}
{which are \textit{reflection positive} (RP).} Here
\begin{equation*}
\mu_{0}^{\Lambda}(d \phi, d \theta):= \prod_{j\in \Lambda} \
\mu_{0}(d \phi_j, d \theta_j)
\end{equation*}
denotes the rotation-invariant product measure, and $\beta$ is the
dimensionless inverse temperature.

{RP is the key property for the \textit{chessboard estimates}. They
allow us then, following e.g. \cite{ES2}, to obtain \textit{contour}
estimates. First we can establish the estimate on the partition
function}
\begin{equation*}
Z^{\Lambda} \geq 1
\end{equation*}
and furthermore, by integrating over intervals $|\theta| \leq
\varepsilon$ and $|\phi| \leq {\pi}/{20}$, we see that also
\begin{equation*}
Z^{\Lambda} \geq (C_{1} \  \varepsilon \exp (2 C_{2} \beta))^{|\Lambda|}
\end{equation*}
with   constants $C_{1}$ and $C_{2}= \cos ({\pi}/{20})$  (which is
close to $1$) which are independent of $\varepsilon$.

On the other hand, let us  call a site \textit{ordered}, if the spin
on that site, as well as all its neighbours, are  in the ditch, and
\textit{disordered}, if it is not in the ditch, and consider the
same universal contour as in \cite{Shl1,ES2}, consisting of
alternating diagonals at distance $2$, which, in turn, consist  of
ordered and disordered sites (separated by sites which are neither);
thus we find that the restricted partition function obtained by
integrating all configurations compatible with the universal contour
satisfies
\begin{equation*}
Z_{univ-cont}^{\Lambda} \leq ((2 \varepsilon)^{{3}/{4}} \exp
(\beta))^{|\Lambda|}.
\end{equation*}
Then,  just as in the proof of Theorem 1 of \cite{ES1}
we obtain
\begin{equation*}
\frac{Z_{univ-cont}^{\Lambda}}{Z^{\Lambda}} \leq
{\varepsilon}^{|\Lambda|/(4+C_{3})}
\end{equation*}
with $C_{3}$ some constant determined by the choice of $C_{1}$ and $C_{2}$.

This implies by standard arguments that, when $\varepsilon$ is
chosen small enough, contours separating ordered and disordered
sites are suppressed, uniformly on a temperature $\Theta$ interval;
since  at low temperatures most sites are ordered and at high
temperatures most sites are disordered, there will be a temperature,
where  disordered and ordered phase(s) (or infinite-volume Gibbs
measures) \textit{coexist}.

We notice that by the Mermin-Wagner theorem
\cite{MerWag,DS2,IofShlVel, Pfi}, in two dimensions all Gibbs
measures are rotation-invariant, so that the spontaneous
magnetisation is necessarily zero. {Since the results of \cite{RZ}
imply that the  generalized XY models (\ref{Ham}) at low
temperatures display a BKT phase, our present statement says  that
for these models with high-exponent-$p$-potentials the transition
between this phase and the high-temperature one is first-order.}

In the three-dimensional version of the model (\ref{Ham}), however, the low-temperature
phase is magnetized, but again the transition in temperature  to the
high-temperature regime is first-order.

Notice also that in general proofs involving contour arguments do not provide very sharp
estimates about the optimal parameter values for which a first-order
transition appears, therefore we will not pursue this road. Below we discuss
what we expect to be the situation based on numerical data, that is,  which is
the value of the parameter $p$ above which one might expect the
first-order transition to appear.

Before ending this section, let us go back in some more detail
to the lattice-gas interpretation of
the square-ditch approximation (\ref{ditch-Ham}).

The theorem just proven implies  that, for $d=2$, the XY lattice-gas
model possesses a first-order transition on  a suitable curve in the
$(\Theta, \varepsilon)$-plane.

This answers an open question from \cite{CSZ} about
the first-order phase transition in $d=2$ diluted plane rotator
model (\ref{ditch-Ham}). On the whole our results  complement those
of \cite{CSZ,GTZ} and \cite{RZ} for this model, since they concern
different parts of the phase diagram in the $(\Theta,\nu)$- or in
the $(\Theta, \varepsilon)$-plane, as well as the mechanism of the
phase transition.

For example, in \cite{GTZ} the existence of the low-temperature BKT
phase in the diluted plane rotator model is proven for a relatively
large \textit{positive} $\nu$,  but without any conclusion on the
mechanism connecting low-temperature and high-temperature behaviour.
On the other hand, the Ginibre and the Wells inequalities applied to
this model (in the same way as it was done for the generalized XY
model in \cite{RZ}) ensure the existence of a low-temperature BKT phase for
\textit{any bounded} $\nu$, but again without conclusion on the
mechanism of the transition.

Moreover, the same analysis leads also to the existence of
a low-temperature BKT phase for generalized XY models
with annealed site-dilution, for all $\nu$, similarly  to what
happens for the plane rotator model.

 In \cite{CSZ} the first-order phase
transition with simultaneous jumps of magnetization and particle
density was established in the $d>2$ diluted plane rotator model in some
domain of both types of $\nu$ (\textit{positive} and  \textit{negative})
at very low temperatures.

In the present note we find (as a byproduct of our generalized-XY-model
analysis) a first order-phase
transition in the diluted plane rotator model (\ref{ditch-Ham}) for
 \textit{moderate negative} $\nu$, since we consider in
(\ref{chem-poten}) sufficiently small $\varepsilon$ as well as
sufficiently small $\Theta$. As mentioned before, this answers positively
the question from \cite{CSZ} about
the first-order phase transition in model (\ref{ditch-Ham}), at least
for those $\nu$.

Based on \cite{CKS}, one knows that  staggered states may
be involved in the mechanism of a first order transition
at positive chemical potentials, at intermediate temperatures;
this will not happen in the regime of negative chemical potentials
which is covered by our results.

\section{Related results and extensions}

Extensions of the above theorem to other pair interaction
potentials are also possible.

For example, the  models studied in \cite{ES1,ES2,ES3,BiKo}
involve $n-$component spins for any ($n \ge 2$) and
possess an $O(n)$ symmetry.  Indeed, their interactions
are functions of the \textit{scalar product} between the two interacting spins,
having the shape of a narrow well.

{The spin interaction in \cite{GZ}, Sect. 2.4, is of the same type
and may be viewed as a diluted version of the Patrascioiu-Seiler model
with a \textit{narrow-ditch interaction}. For discussion of its
low-temperature phase see e.g. \cite{Aiz}.}

The proof scheme indicated above  works for various
\textit{combinations} of \textit{single-site} terms and
\textit{nearest neighbour} interaction terms, of which at least some
need to have a narrow shape. The spins can be $n$-component spins,
and the symmetry can be either O(n) or O(2) or some symmetry in
between.
 {For example, we might  have  narrow-ditch single-site
potentials (as mentioned above), or narrow-well single-site
potentials as in \cite{DS1}, as well as narrow-well interactions
(\cite{ES1,ES2,ES3}) or narrow-ditch interactions (see \cite{GZ},
Sect. 2.4, and \cite{BiChStar})}. The interactions can have
\textit{one} well (or ditch) for ferromagnets, \textit{two} (for
liquid crystal models, possessing $RP^{(n-1)}$ symmetry), or more
and could also include diagonal nearest neigbour terms (as in
\cite{MeNa}, inspired by the model of \cite{Shl0}). The narrowness of
such terms then either creates or reinforces the first-order
behaviour.

{Another kind of possible extensions is related to \textit{quantum}
versions of our models. This observation is inspired by the recent
paper \cite{BiChStar}, which studies, in particular, a non-linear
quantum XY model. The main ingredient for their arguments is the
\textit{quantum} RP property, which is a quite subtle matter, but
the \textit{ferromagnetic} quantum XY model does verify it. Since in
(\ref{ditch-Ham}) the interaction terms are multiplied by  simple
single-site classical random variables (the scalar occupation
numbers), the square-ditch Hamiltonian also verifies the
\textit{quantum} RP property. Then according to \cite{BiChStar} we
can claim the existence of the first order phase transition in this
quantum model, since we proved it for the classical model
(\ref{ditch-Ham}) and we know that its quantum analogue verifies the
quantum RP property.}

\section{Numerical estimates of transition orders and temperatures}

When $d=3$, a Mean Field (MF) study of the ordering transition is at
least qualitatively correct, and relatively feasible in
computational terms; moreover,  this treatment can be refined by
using various cluster-variational techniques; we used here a
Two-Site Cluster (TSC) approach, and both treatments follow Ref.
\cite{RZ}. Calculations were carried out for $5 \le p \le 12$, and
then $p=16,20$.
\\
In both cases we found that, upon increasing $p$, the transition
changes from second to first order;  the two treatments exhibited
different thresholds, i.e. a threshold of $p$ between $5$ and $6$
for MF,  and a threshold of $p$ between $10$ and $11$ for TSC;
results of both treatments are presented  and compared in Tables
(\ref{t01}) and (\ref{t02}), where first-order transitional
properties, such as the energy jump, $\Delta U^*$, and the order
parameter at the transition, $\overline{M}$, are shown for $p \ge
11$, where both treatments predict a first-order transition.

As a side remark, we also notice that the results of \cite{BiCh}
imply the existence of a first-order transition for any $p \ge 6$ in
sufficiently high dimension.

On the other hand, simulation results, obtained  for $d=3$
and to be reported elsewhere \cite{CR},
show a second-order transition for $p=8$, and suggest a
first-order one for $p=12,16,20$.%

\bigskip

\noindent
\section*{Acknowledgements}
A.v.E. and V.A.Z. thank Senya Shlosman for the very enjoyable
earlier collaborations on reflection positive models, which directly
led to the proof given above. We also thank Christof K\"ulske for a helpful
remark.

\newpage
\parindent 0cm
\begin{table}[ht!]
\caption[]{
MF  results for  transitional properties
of the generalized XY models in three dimensions, obtained
with different values of the exponent  p.
\\~~~}
\label{t01}
\begin{tabular}{rlccc}
\hline
\\
p & $\Theta_{MF}$ & type & $\Delta U^*$ & $\overline{M}$
\\
\hline
\\
      5 & ~1.1082~     & II & ~ & ~
      \\
      6 &  ~1.0287~ & I & ~ & ~
      \\
      7 &  ~0.9741~ & I & ~ & ~
      \\
      8 &  ~0.9336~ & I & ~ & ~
      \\
      9 &  ~0.9019~ & I & ~ & ~
      \\
      10 &  ~0.8762~ & I & ~ & ~
      \\
      11 &  ~0.8548~ & I &  ~1.2336~ &  ~0.7506
      \\
      12  &  ~0.8366~ & I&   ~1.3140~ &  ~0.7687
      \\
      16 &  ~0.7836~ & I &  ~1.5355~ &  ~0.7836
      \\
      20 & ~0.7486~ & I & ~1.6712~  &  ~0.8387
\\
~ & ~ & ~ & ~ & ~
\\
\hline
\end{tabular}
\end{table}
\begin{table}[h!]
\caption[]{
TSC  results for
transitional properties
of the generalized XY models in three dimensions, obtained
with different values of the exponent  p.
\\~~~}
\label{t02}
\begin{tabular}{rlccc}
\hline
\\
p & $\Theta_{TSC}$ & type & $\Delta U^*$ & $\overline{M}$
\\
\hline

\\
 5 &     ~1.1011~ & II & ~ & ~
 \\
   6    &  ~1.0416~ & II & ~ &
 \\
 7  &   ~0.9935~ & II & ~ & ~
 \\
 8 &  ~0.9537~ & II & ~ & ~
 \\
  9 &   ~0.9199~ & II &  ~ & ~
 \\
    10 &   ~0.8907~ & II & ~ & ~
 \\
    11 &   ~0.8659~  & I & ~0.3242~ &  ~0.3994~
 \\
    12  &   ~0.8461~ & I & ~0.5437~ &  ~0.5098~
 \\
    16  &   ~0.7906~ & I & ~1.0097~ &  ~0.6721~
 \\
    20  &   ~0.7549~ & I & ~1.2578~ &  ~0.7374~
\\
~ & ~ & ~ & ~ & ~
\\
\hline
\end{tabular}
\end{table}

\addcontentsline{toc}{section}{\bf References}


\begin{thebibliography}{99}

\bibitem{Aiz} M.~Aizenman.
\newblock On the slow decay of $O(2)$ correlations in the absence of topological excitations:
remark on the Patrascioiu-Seiler model.
\newblock {\em J.Stat.Phys.}, 77:351--359, 1994.

\bibitem{BiChStar} M.~Biskup, L.~Chayes and S.~Starr.
\newblock Quantum spin systems at finite temperature.
\newblock arXiv:math-ph{/}0509017, to appear in {\em Comm.Math.Phys.},
2006.

\bibitem{BiCh} M.~Biskup and L.~Chayes.
\newblock Rigorous analysis of discontinuous phase transitions via mean-field bounds.
\newblock {\em Commun.Math.Phys.}, 238:53--93, 2003.

\bibitem{BiKo} M.~Biskup and R.~Koteck\'y.
\newblock Forbidden gap argument for
phase transitions proved by means of chessboard estimates.
\newblock {\em Commun. Math. Phys.}, 264:631-656, 2006.

\bibitem{CR} H.~Chamati and S.~Romano.
\newblock Article in preparation.

\bibitem{CKS}L.~Chayes, R.~Koteck\'y and  S.~B.~Shlosman.
\newblock Staggered phases in diluted systems with continuous spins.
\newblock {\em Commun.Math.Phys.}, 189:631--640, 1997.

\bibitem{CSZ} L.~Chayes, S.~B.~Shlosman and  V.~Zagrebnov.
\newblock Discontinuity of the Magnetization in Diluted $O(n)$-Models.
\newblock \textit{J.Stat.Phys.},  98: 537--549,  2000.

\bibitem{CRMP}
H.~Chamati, S.~Romano, L.~A.~S.~M\'ol and A.~Pereira.
\newblock Three-dimensional generalized XY models: A Monte Carlo study.
\newblock {\em Europhysics Lett.}, 72:62--69, 2005.

\bibitem{DS1} R.~L.~Dobrushin and S.~B.~ Shlosman. \newblock Phases
corresponding to the local minima of the energy. \newblock{\em
Selecta Math. Soviet.}, 1: 317--338, 1981.

\bibitem{DS2} R.~L.~Dobrushin and S.~B.~ Shlosman. \newblock Absence of
Breakdown of Continuous Symmetry in 2-dimensional Models of Statistical
Physics. \newblock {\em Commun.Math.Phys.}, 42:31--40, 1975.

\bibitem {EFS}A.~C.~D. van Enter, R.~Fern{\'a}ndez, and A.~D. Sokal.
\newblock Regularity properties and pathologies of position-space
renormalization-group transformations: Scope and limitations of
{G}ibbsian theory.
\newblock{\em J. Stat. Phys.}, 72:879--1167, 1993.

\bibitem{ES1}A.~C.~D. van Enter and S.~B.~Shlosman. \newblock First-Order
Transitions for n-Vector models in Two and More Dimensions: Rigorous Proof.
\newblock{\em Phys. Rev. Lett.}, 89:285702, 2002.

\bibitem{ES2}A.~C.~D. van Enter and S.~B.~Shlosman. \newblock Provable
First-Order Transitions for Nonlinear Vector and Gauge
Models with Continuous Symmetries.
\newblock {\em Commun.Math.Phys.}, 255:21--32, 2005.

\bibitem{ES3} A.~C.~D. van Enter and S.~B.~Shlosman. \newblock First-order
transitions for very nonlinear sigma models.
\newblock Arxiv cond-mat{/}0506730, {\em Markov Proc.Rel.Fields},to appear,
2006.

\bibitem{FL} J.~Fr\"ohlich and E.~H.~Lieb. \newblock Phase Transitions in
Anisotropic Lattice Spin Systems.
\newblock{\em Commun.Math.Phys.}, 60:233--267, 1978.

\bibitem {Geo}H.-O.~Georgii. \newblock{\em
Gibbs Measures and Phase Transitions}. \newblock Walter de Gruyter (de Gruyter
Studies in Mathematics, Vol.\ 9), Berlin--New York, 1988.

\bibitem {GeoHagMae}H.-O.~Georgii, O.~H{\"a}ggstr{\"o}m and C.~Maes. \newblock
The random geometry of equilibrium phases. \newblock Phase transitions and
critical phenomena (C. Domb and J.L. Lebowitz, Eds.), vol. 18,
Academic Press, London, 2001.

\bibitem {GZ} H.-O.~Georgii and V.~Zagrebnov.
\newblock Entropy-Driven Phase Transitions in Multitype Lattice Gas Models,
\newblock \textit{J.Stat.Phys.}, 102: 35--67,  2001.

\bibitem {GTZ} C.~Gruber, H.~Tamura and V.~Zagrebnov.
\newblock Berezinski\v\i-Kosterlitz-Thouless Order in Two-Dimensional
$O(2)$-Ferrofluid,
\newblock \textit{J.Stat.Phys.} 106: 875-893,  2002.

\bibitem{IofShlVel} D.~Ioffe, S.~B.~Shlosman and Y.~Velenik.
\newblock 2D Models of Statistical Physics with Continuous Symmetry:
The Case of Singular Interactions.
\newblock{\em Commun. Math. Phys.}, 226: 433--454, 2002.

\bibitem{KS}R.~Koteck\'y and S.~B.~Shlosman. \newblock First-order
transitions in large entropy lattice models.
\newblock{\em Commun.Math. Phys.}, 83:493--515, 1982.

\bibitem{MeNa} A.~Messager and B.~Nachtergaele. \newblock A Model with
Simultaneous First and Second Order Phase Tranditions.
Arxiv cond-mat{/}0501229, {\em J.Stat.Phys.}, 122:1--14, 2006 .

\bibitem {MerWag}N.~D.~Mermin and H.~Wagner. \newblock Absence of
Ferromagnetism or Antiferromagnetism in One- or Two-Dimensional Heisenberg
Models. \newblock{\em Phys. Rev. Lett.}, 17:1133--136, 1966.


\bibitem{MPCR}
L.~A.~S.~M\'ol, A.~R.~Pereira, H.~Chamati and S.~Romano.
\newblock Monte Carlo study of generalized XY models.
{\em Eur. J. Phys} {\bf B}, to appear, 2006.

\bibitem{MPMM}
L.~A.~S.~M\'ol, A.~R.~Pereira and W.~A.~Moura-Melo.
\newblock On phase transition and vortex stability in the generalized
XY models.
\newblock {\em Phys. Lett.} {\bf A}, 319:114--121, 2003.

\bibitem{Pfi} C.-E. Pfister. \newblock On the Symmetry of the Gibbs
States in Two-dimensional Lattice Systems. \newblock{\em Commun. Math. Phys.},
79:181--188, 1981.

\bibitem {RP}The method of Reflection Positivity was developed by F.~J.~Dyson,
J.~Fr\"ohlich, R.~B.~Israel, E.~H.~Lieb, B.~Simon and T.~S.~Spencer in a
series of papers. It is described in the last chapters of \cite{Geo} or in
\cite{Shl1}. The chessboard estimate method goes back to \cite{FL}.

\bibitem{RZ}
S.~Romano and V.~Zagrebnov.
\newblock On the XY model and its generalizations.
\newblock {\em Phys. Lett.} {\bf A}, 301:402--407, 2002.

\bibitem{Shl0} S.~B.~Shlosman.
\newblock  Phase transitions for two-dimensional models with isotropic short-range interactions
and continuous symmetries.
\newblock{\em Commun. Math.Phys.}, 71: 207--212, 1980.

\bibitem {Shl1}S.~B.~Shlosman. \newblock The method of reflection positivity in
the mathematical theory of first-order phase transitions. \newblock{\em
Russian Mathematical Surveys}, 41: 83--145, 1986.

\bibitem{Sim} B.~Simon. The Statistical Mechanics of Lattice Gases, Vol 1.
Princeton University Press, Princeton N.J., 1993.

\bibitem{Sin}Ya.~G.~Sinai. Theory of Phase Transitions: Rigorous Results,
Pergamon Press, Budapest, 1982.

\end{thebibliography}
\end{document}